\documentclass[usenatbib]{mnras}
\usepackage{graphicx}
\usepackage{amsmath}
\usepackage{amssymb}
\usepackage[utf8]{inputenc}

\newcommand\reffig[1]{Figure \ref{#1}}
\newcommand\refeq[1]{eq.~(\ref{#1})}

\title{Quantifying the power spectrum of small-scale structure in semi-analytic galaxies}

\author[Brennan et al.]{\parbox{\textwidth}{
Sean Brennan,$^{1}$
Andrew J. Benson,$^{2}$
Francis-Yan Cyr-Racine,$^{3}$
Charles R. Keeton,$^{1}$
Leonidas A. Moustakas,$^{4,5}$
and Anthony R. Pullen$^{6}$
}\\
\\
\parbox{\textwidth}{
$^{1}$Department of Physics and Astronomy, Rutgers,
The State University of New Jersey, Piscataway, New Jersey 08854, USA \\
$^{2}$Carnegie Observatories, 813 Santa Barbara Street, Pasadena, CA 91101, USA \\
$^{3}$Department of Physics, Harvard University, Cambridge, Massachusetts 02138, USA \\
$^{4}$NASA Jet Propulsion Laboratory, California Institute of Technology, Pasadena, California 91109, USA \\
$^{5}$California Institute of Technology, Pasadena, California 91125, USA \\
$^{6}$Center for Cosmology and Particle Physics, Department of Physics, New York University, 726 Broadway, New York, NY, 10003, USA
}}

\begin{document}
\label{firstpage}
\pagerange{\pageref{firstpage}--\pageref{lastpage}}
\maketitle

\begin{abstract}
In the cold dark matter (CDM) picture of structure formation, galaxy mass distributions are predicted to have a considerable amount of structure on small scales. Strong gravitational lensing has proven to be a useful tool for studying this small-scale structure. Much of the attention has been given to detecting individual dark matter subhalos through lens modeling, but recent work has suggested that the full population of subhalos could be probed using a power spectrum analysis. In this paper we quantify the power spectrum of small-scale structure in simulated galaxies, with the goal of understanding theoretical predictions and setting the stage for using measurements of the power spectrum to test dark matter models. We use a sample of simulated galaxies generated from the \texttt{Galacticus} semi-analytic model to determine the power spectrum distribution first in the CDM paradigm and then in a warm dark matter scenario. We find that a measurement of the slope and amplitude of the power spectrum on galaxy strong lensing scales ($k\sim 1$ kpc$^{-1}$) could be used to distinguish between CDM and alternate dark matter models, especially if the most massive subhalos can be directly detected via gravitational imaging.  
\end{abstract}


\section{Introduction}
\label{sec:Intro}

Dark matter is a key component of the standard model of cosmology, but its fundamental nature remains uncertain. In the standard Cold Dark Matter (CDM) model of cosmological evolution, structures form through the accretion and merging of smaller structures. This bottom-up picture of structure formation leads to dark matter halos that contain substructure in the form of smaller, less massive subhalos. Cosmological simulations make specific predictions about the mass function and spatial distributions of this dark matter substructure \citep[e.g.,][]{springel, boylan, ponos_proj}. These predictions depend strongly on the type of dark matter particle considered. For instance, moving from CDM to a warm dark matter (WDM) model by decreasing the mass of the dark matter particle reduces the amount of substructure in galaxies \citep[e.g.,][]{wdm_goetz,Lovell:2013ola,Bose:2016irl}. This difference provides a possible way to learn about the fundamental nature of dark matter by observing the abundance of satellite galaxies within the Local Group \cite[see, e.g.,][]{Anderhalden:2012jc,2015MNRAS.448..792G,Schneider:2014rda}. 

In practice, the actual number of small dwarf galaxies surrounding the Milky Way depends not only on the dark matter physics, but also on the star formation efficiency in small dark matter subhalos \citep[e.g.,][]{Bullock,Benson02,Somerville,Behroozi:2012iw,Garrison-Kimmel:2013eoa,Brooks,Brook:2013laa,2017MNRAS.470..651R}. While there are still considerable uncertainties in the stellar content of small halos, it appears plausible that dark matter halos below a certain mass threshold may be entirely devoid of stars \cite[see, e.g.,][]{Dooley:2016xkj,Kim:2017iwr}. Therefore, directly observing the substructure content of the Local Group at the smallest scales is very challenging, although indirect methods based on the gravitational influence of small subhalos on the Milky Way disk \citep{Feldmann:2013hqa}, halo stars \citep{Buschmann:2017ams}, or stellar streams \citep{2014ApJ...788..181N,2016MNRAS.463..102E,2016ApJ...820...45C,2016PhRvL.116l1301B,2017MNRAS.466..628B,Banik:2018pjp} could potentially shed light on local small-scale structure.

Since it is sensitive to the total projected mass distribution along the line of sight between the high-redshift source and the observer, gravitational lensing provides a means for detecting dark matter subhalos even if they do not contain any stars or gas. While the technique could in principle be applied to our local neighborhood \citep[see, e.g.,][]{Erickcek:2010fc,VanTilburg:2018ykj}, gravitational lensing is the only way to detect dark substructure in cosmologically distant galaxies. In observed gravitational lenses, substructure appears as localized perturbations to an otherwise ``smooth'' mass model responsible for setting the broad structure of the lensed images. These perturbations are usually detected through anomalies in the lensing observables that cannot be easily reabsorbed by a change to the smooth lens model \citep{kgp_cusp, kgp_fold,Koopmans:aa,Vegetti:2008aa,Hezaveh:2012ai}. In some cases, these anomalies can be well fit by the inclusion of a mass clump in the model. This is often interpreted as evidence of the ability to detect individual dark matter subhalos with gravitational lensing \citep{Mao98,metcalf01,Vegetti_2010_1,Vegetti_2010_2,vegetti2012,Nierenberg:2014aa,2014MNRAS.442.2017V,hez_clump}. We note, though, that translating a substructure detection to the actual physical properties of a dark matter subhalo has important subtleties \citep{Minor:2016jou,Daylan:2017kfh}. Also, some of these anomalies could be caused by baryonic substructure, although it is statistically unlikely that all of the observed anomalies are caused by baryons \citep{Hsueh_16, Hsueh_17,Hsueh_18,gilman}.

CDM theory predicts the existence of abundant small-scale structure and so it would be convenient to build inference models that are able to capture the collective effect of this substructure. There has been work done to this end that has incorporated a population of subhalos within lens models in a statistical way \citep{dalal_kochanek,fadely,Birrer2017}. Work has also been done to calculate what effect a population of subhalos can have on the image positions and relative time delay of multiply-imaged quasars \citep{FYCR}.

Another way to capture the statistical properties of the small-scale structure within lens galaxies is with a power spectrum analysis. It has previously been shown that measuring the power spectrum of projected density fluctuations with current observations of strongly-lensed images is likely feasible \citep{hez_pow,Chatterjee,Bayer,Pksub}. Moreover, theoretical predictions for the shape and amplitude of the substructure convergence power spectrum from realistic populations of subhalos has recently been presented in \cite{Diaz}. There, it was shown that the substructure power spectrum contains important information about the abundance, masses, and density profiles of the subhalos inhabiting the lens galaxy.

Substructure lensing is moving towards analyses that include a power spectrum piece that accounts for small-scale structure of the kind predicted by current dark matter theories. In order for measurements of the power spectrum to be useful for weighing competing theories of dark matter we must first determine what these theories look like in the language of power spectra. In this work we move beyond theoretical estimates to directly quantify the lensing convergence power spectrum in simulated galaxies with an eye towards informing future lensing measurements and with the hope that the power spectrum formalism becomes the new standard for analyzing the substructure content of lens galaxies.

This paper is organized as follows. In Section \ref{sec:Methods} we describe our subhalo populations and outline the method for calculating the substructure power spectrum. In Section \ref{sec:Results-CDM} we present the results of our calculation of the power spectrum distribution for our CDM populations and show how it is affected by removing massive subhalos. We also test the validity of using multiple projections of individual subhalo populations as a proxy for having independent populations. Finally, in Section \ref{sec:Results-WDM} we compare our CDM and WDM subhalo populations in terms of their power spectrum distributions.

\section{Methods}
\label{sec:Methods}

\subsection{Subhalo Populations}

\begin{figure}
\includegraphics[width=\linewidth]{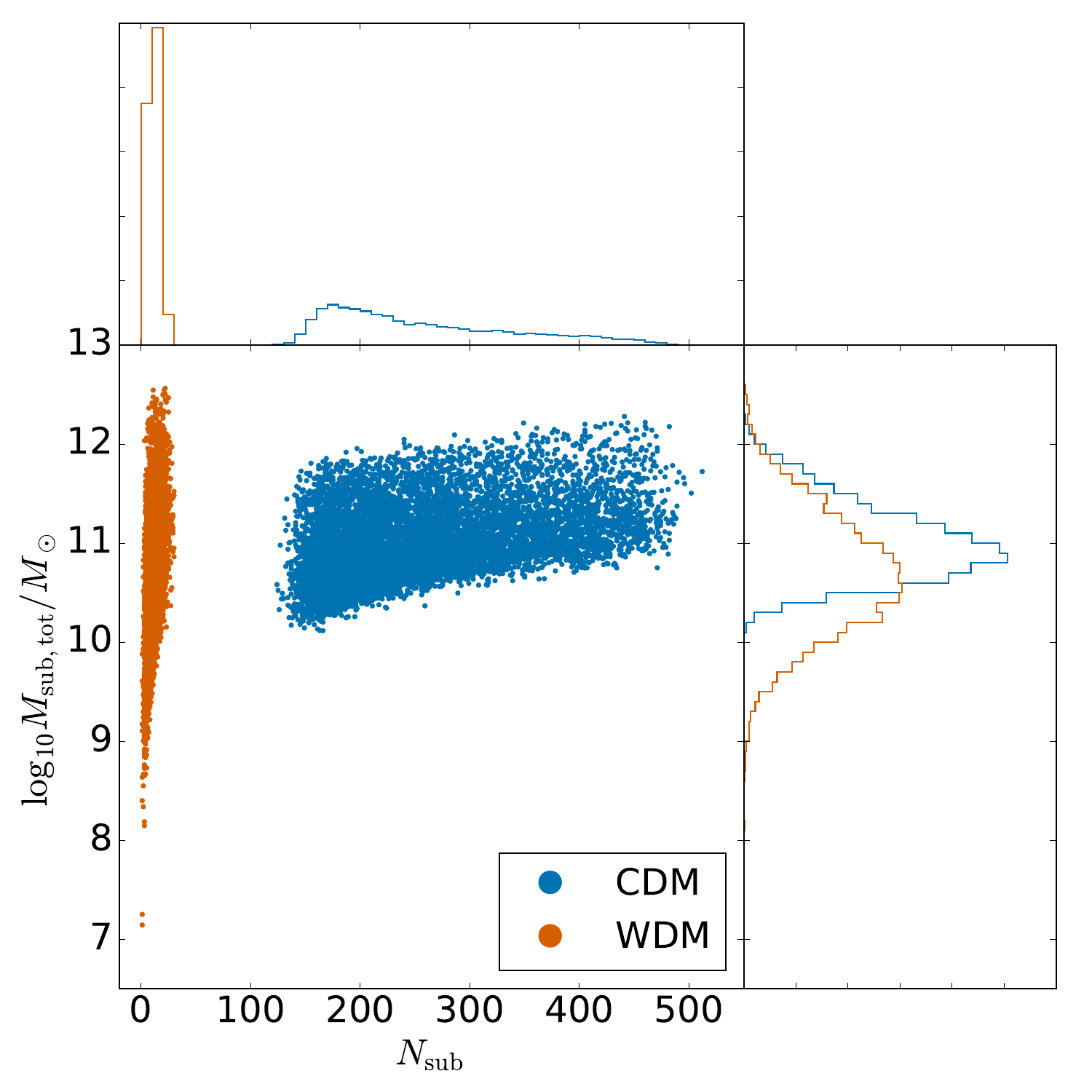}
\caption{The total mass in subhalos $\log_{10}M_{\mathrm{sub,tot}}$ vs. number of subhalos $N_{\mathrm{sub}}$ for 9160 subhalo populations. CDM populations are shown in blue and WDM populations in orange. Side panels show 1D histograms for $\log_{10}M_{\mathrm{sub,tot}}$ and $N_{\mathrm{sub}}$. }
\label{N_M}
\end{figure}

We use the semi-analytic galaxy-formation model \texttt{Galacticus} \citep{galacticus} to generate the galactic dark matter halos used in our analysis. \texttt{Galacticus} assumes that the host halo and subhalos are spherical and that the subhalo population is statistically isotropic. Our \texttt{Galacticus} simulations are the same set used for the mass function analysis by \cite{pullen}. In their paper they show that results from \texttt{Galacticus} agree broadly with those from $N$-body simulations; specifically, they compare the differential subhalo abundance and the cumulative subhalo abundance with the Aquarius simulations and the Via Lactea II simulation, respectively. The simulations we use contain only dark matter and include the effects of tidal heating, tidal stripping, and dynamical friction. We have two sets of 9160 halos: one uses the standard CDM model of dark matter physics and the other is a WDM model with a dark matter particle mass of $1.5$ keV. While this choice of mass is technically ruled out by observations of the Lyman-$\alpha$ forest \citep{Yeche:2017upn,Irsic:2017ixq} and the abundance of local satellite dwarf galaxies \citep{Schneider:2014rda,Escudero:2018thh}, it does provide us with a model that is significantly different than CDM, hence making it easier to highlight the differences between the two dark matter candidates. Main halo masses range from $1-3 \times 10^{12} M_{\odot}$ and the mass resolution of the simulation is $M_{\rm res} = 5 \times 10^7 M_{\odot}$. The main halo is removed from the mass model (hence leaving only the subhalo population) before computing the substructure power spectrum. Here we focus exclusively on the subhalo contribution to the power spectrum, and leave to future work the study of its line-of-sight contribution \citep[see, e.g.,][]{Keeton:2003aa,Despali:2017ksx, gilman2019}. 

Our simulated halos are on the low end of the mass range typically probed by galaxy-scale strong lensing. Also, they have been evolved to redshift $z=0$, whereas most lens galaxies are at redshifts between about 0.2 and 0.8. For both of these reasons, our simulated halos are likely to have less substructure than might be expected in typical lens galaxies \citep[see][]{z_evol}. As such, the substructure power spectra presented in this work should be taken as conservative lower limits on their possible amplitude. Importantly, the simulations provide a sample that is large enough to characterize the statistical variability from one lens to the next.

\reffig{N_M} shows the distribution of the number of subhalos $N_{\mathrm{sub}}$ vs. the total mass in subhalos $M_{\mathrm{sub}}$ for all 9160 subhalo populations. The CDM populations have an average of $\langle N_{\mathrm{CDM}}\rangle = 257$ subhalos and the WDM populations have an average of $\langle N_{\mathrm{WDM}}\rangle = 11$ subhalos.

Each subhalo is parametrized as a truncated NFW halo with 3-d density profile
\begin{equation}
\rho (r) = \frac{M_0}{4\pi r(r+r_s)^2}\left(\frac{r_t^2}{r^2+r_t^2}\right)
\label{eq:rho}
\end{equation}
where $r_s$ is the scale radius and $r_t$ is the tidal truncation radius. The lensing properties of truncated NFW halos are given by \citet{baltz}.

\subsection{Convergence Maps and Power Spectra}

\begin{figure}
\includegraphics[width=\linewidth]{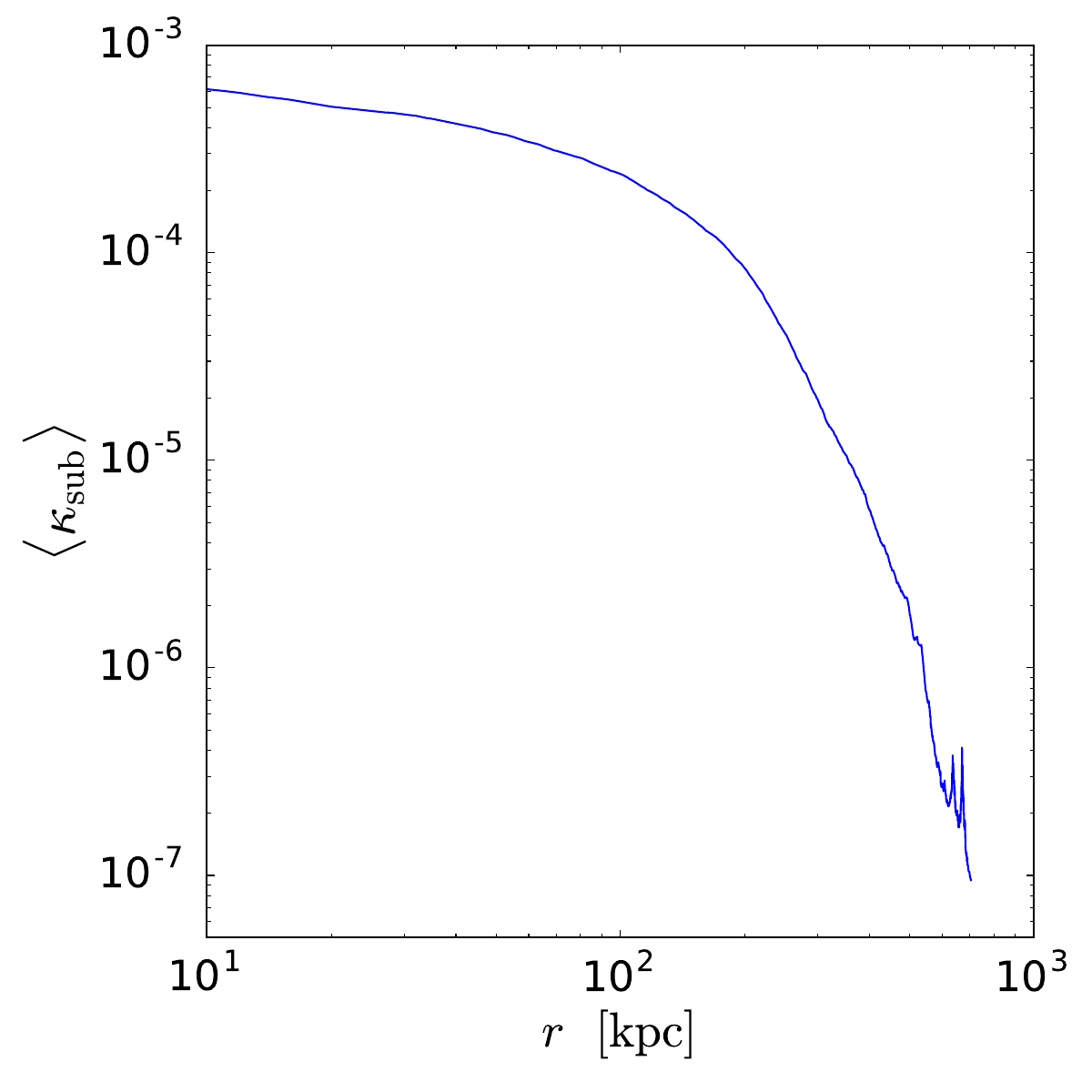}
\caption{Mean convergence profile for CDM, after taking both an ensemble average and a circular average.
}
\label{kapr}
\end{figure}

For the single-plane lensing we consider here, the quantity of interest is the projected surface mass density scaled by the critical density for lensing.
We use a critical density of $\Sigma_{\mathrm{crit}} = 1.15 \times 10^{11} M_\odot/\mbox{arcsec}^2$ throughout this work. This critical density could be realized for a system with lens redshift $z_l = 0.5$ and source redshift $z_s = 1.0$. For a lens at $z_l = 0.5$, 1 arcsecond  corresponds to 6.1 kpc. The maps we use are $1000 \times 1000$ pixels corresponding to $\sim 1.2$ Mpc on a side. In \reffig{kapr} we show an ensemble average of the convergence in substructure, $\langle \kappa_{\mathrm{sub}} \rangle$, vs.\ radius $r$. We can see that the overall convergence in substructure is relatively uniform in the inner $\sim 100$ kpc and has a typical value of $\langle \kappa_{\mathrm{sub}}\rangle \sim 6 \times 10^{-4}$.

\begin{figure*}
\includegraphics[width=\textwidth]{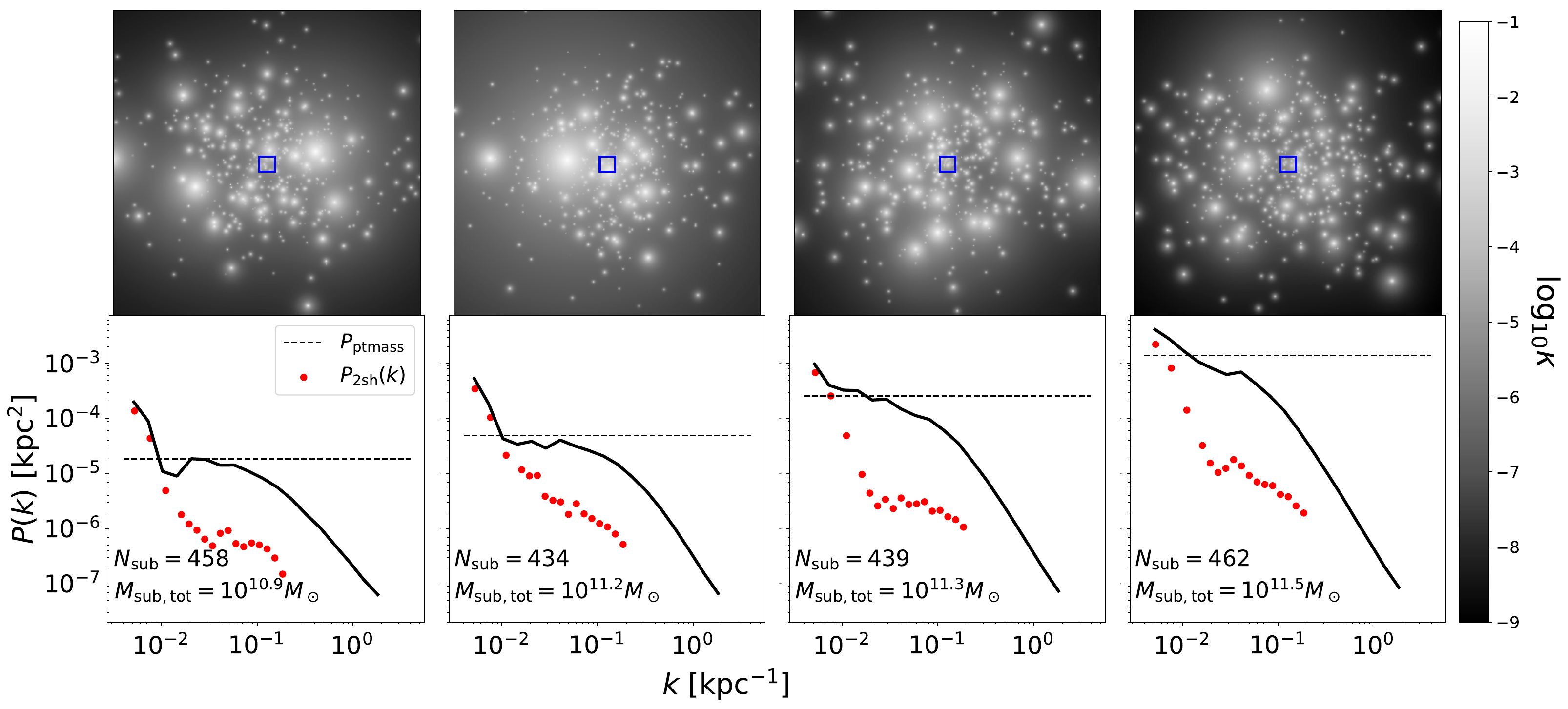}
\caption{
\emph{Top:} Convergence maps for CDM substructure populations chosen to reflect the range of low-$k$ power present in our simulations. The grayscale indicates the convergence $\kappa = \Sigma/\Sigma_{\rm crit}$. Each bright spot is a dark matter subhalo.  The full box size is $1.2 \times 1.2$ Mpc$^2$. The blue squares measure $60 \times 60$ kpc$^2$ and mark the smaller boxes used in Section 3.
\emph{Bottom:} 1-d power spectra corresponding to the maps in the top panels. Each panel includes a label indicating the number of subhalos in the population, $N_{\rm sub}$, and their total mass, $M_{\rm sub,tot}$. The dashed lines show estimates of the power treating all subhalos as point masses, using \refeq{eq:Pptmass}. The red points show the 2-subhalo term, $P_{\rm 2sh}(k)$.
}\label{kapmaps}
\end{figure*}

\reffig{kapmaps} shows example of individual convergence maps for CDM. For each map, we compute the 2-d Fourier transform and square it to get a map of the power. We then take a circular average to obtain the 1-d power spectrum. Examples of individual power spectra are shown in \reffig{kapmaps}.

These power spectra have four characteristic features: a normalization, an upturn at low-$k$, a turnover scale, and a high-$k$ slope. The physical origin of these features are discussed by \citet{Diaz}. Briefly, the normalization is determined by the overall convergence in substructure: $P \propto \langle \kappa_{\rm sub} \rangle \langle M^2 \rangle/(\langle M \rangle \Sigma_{\rm crit})$ where $\langle M \rangle$ is the average subhalo mass, $\langle M^2 \rangle$ is the second moment of the subhalo mass function, and the remaining proportionality factor involves the internal structure of the subhalos. A simple estimate of the normalization can be made if we approximate the subhalos as point masses:
\begin{equation}
  P_{\rm ptmass} = \frac{1}{A} \sum_{i=1}^{N}{m^2}
\label{eq:Pptmass}
\end{equation}
where $m = M/\Sigma_{\rm crit}$ is a normalized mass that has dimensions of area, and $A$ is the area of the convergence map. \reffig{kapmaps} includes the point mass power estimate as dashed lines.

The upturn visible at low-$k$, especially in the first two power spectra of \reffig{kapmaps}, primarily comes from the nonuniform spatial distribution of subhalos. This feature is imprinted on the subhalo population by the host halo and is called the 2-subhalo term, $P_{\rm 2sh}(k)$. We plot $P_{\rm 2sh}(k)$ for the individual populations as red points in \reffig{kapmaps}. The 2-subhalo term is only important at the smallest $k$ and rapidly becomes subdominant compared to the 1-subhalo contribution as the wavenumber is increased.

Finally, the turnover at $k \approx 0.1$ kpc$^{-1}$ is related to the truncation radii of the subhalos, and the high-$k$ slope is determined by the choice of density profile.

\section{Results for CDM}
\label{sec:Results-CDM}

\subsection{Full Subhalo Population}

\begin{figure*}
\centering
\includegraphics[width=\textwidth]{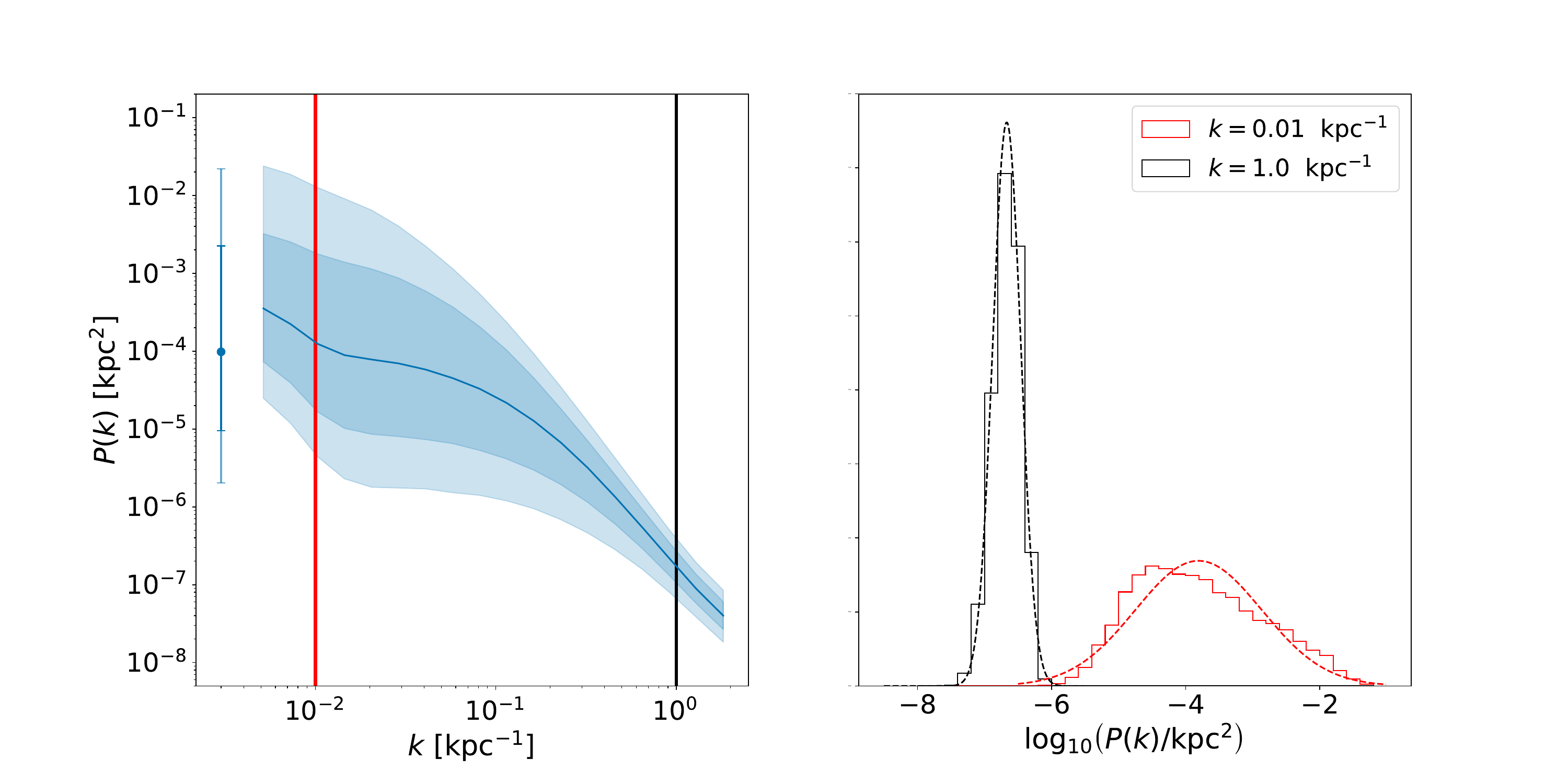}
\caption{\emph{Left:} The convergence power spectrum distribution for CDM. The solid line is the median and the shaded bands show the 68\% and 95\% confidence regions. The vertical lines indicate the $k$ values used in the right panel. The point with error bars to the left of the distribution shows the median, 68\%, and 95\% confidence intervals of the point mass power distribution.
\emph{Right:} Slices of $\log_{10}(P)$ at $k = 0.01$ kpc$^{-1}$ and $k = 1.0$ kpc$^{-1}$. The dashed lines are Gaussians with the same mean and standard deviation as the distributions ($\mu = -6.8$ and $\sigma = 0.2$ for the black curve, and $\mu = -4.0$ and $\sigma = 1.0$ for the red curve).}
\label{pspec_cdm}
\end{figure*}

We repeat the procedure outlined in Section 2.2 for our 9160 CDM subhalo populations. 

The resulting power spectrum distribution is shown in the left panel of \reffig{pspec_cdm}. The overall shape of the full distribution is the same as the individual power spectra shown in \reffig{kapmaps}. There is approximately an order of magnitude scatter that reflects the map-to-map variations. The point with error bars to the left of the distribution shows the median, 68\%, and 95\% confidence intervals of the point mass power (Eq.~\ref{eq:Pptmass}). The fact that the variance in point mass power closely matches the spread in the power spectrum distribution at low-$k$ indicates that the scatter in our distribution is mainly due to differences in the subhalo abundance between populations. 

At high-$k$ the scatter is reduced due to the similarity of our maps at small spatial scales. This is because the subhalos in our populations have a fixed density profile. The right panel of \reffig{pspec_cdm} shows that the distribution of power at fixed $k$ is approximately log-normal at both low and high values of $k$.

\subsection{Impact of Most Massive Subhalos}

\begin{figure}
\includegraphics[width=\linewidth]{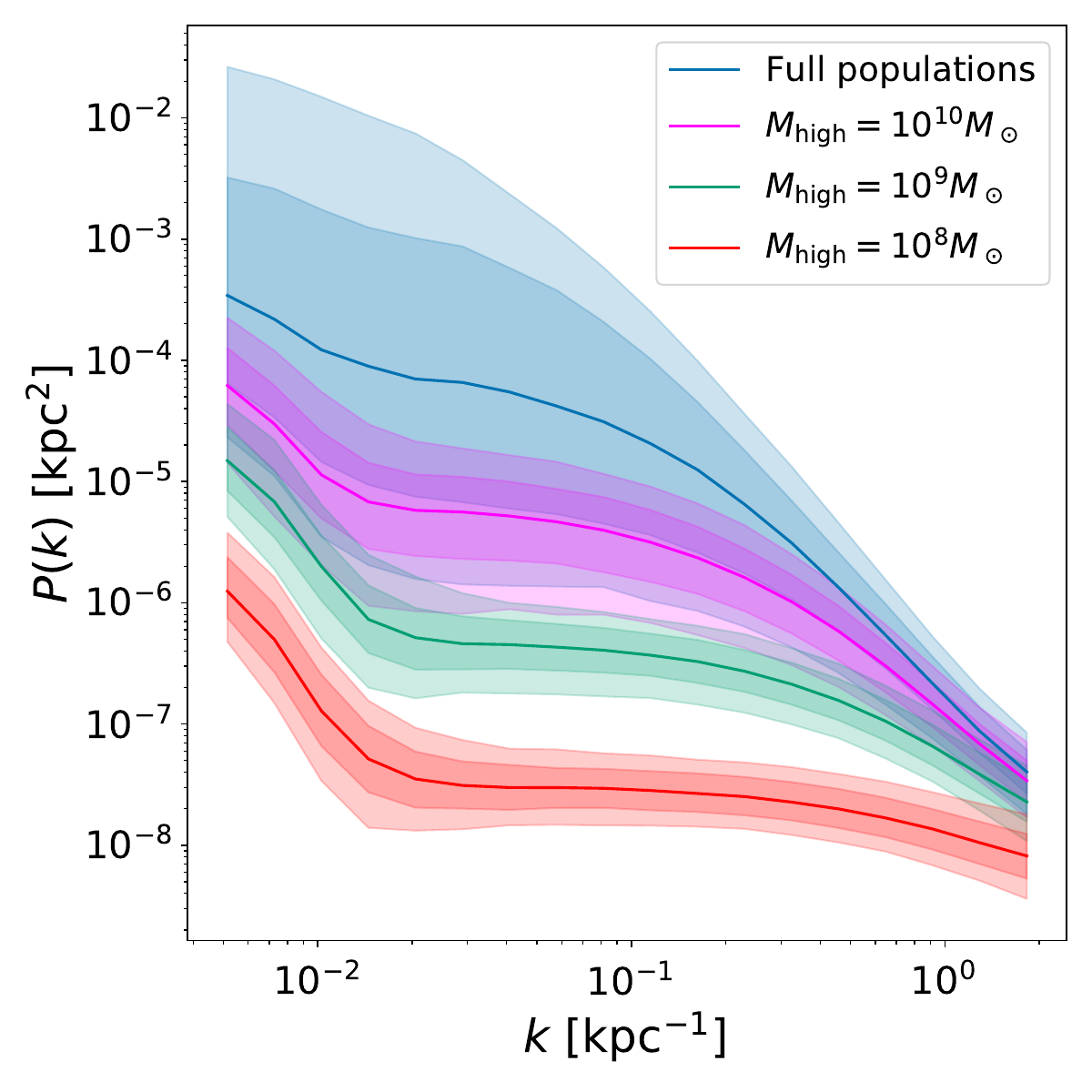}
\caption{Comparison of the CDM power spectrum distribution for populations with different choices of highest subhalo mass. Colors correspond to different values of $M_{\rm high}$. Again, the solid lines are the median values and the bands are 68\% and 95\% confidence. The full populations contain an average of 257 subhalos. The sub-populations contain averages of 255, 246, and 186 subhalos for highest mass of $10^{10} M_{\odot}$, $10^{9} M_{\odot}$, and $10^{8} M_{\odot}$ respectively.}
\label{cdm_cuts}
\end{figure}

We now seek to understand the physical origin of the large scatter in \reffig{pspec_cdm}. We note that massive subhalos have large contributions to the power but are statistically rare (compare the different panels in \reffig{kapmaps}), so they may not be suitable for a power spectrum treatment. Furthermore, their effects on lensed images may be non-perturbative, so they might have to be explicitly incorporated into lens models. Motivated by these ideas we introduce a mass threshold and remove subhalos above $M_{\rm high}$ before computing the power spectrum. \reffig{cdm_cuts} shows how removing the most massive subhalos affects the power spectrum distribution for CDM.

With $M_{\rm high} = 10^{10} M_\odot$ only the 1--2 most massive subhalos are removed on average. It is not surprising that removing these subhalos reduces the overall power, but it is striking how much it decreases the variance in the power spectrum distribution. A large portion of the variance apparently arises from the most massive subhalos because they are rare, and statistical variations lead to large difference in the overall power.

Decreasing the largest subhalo mass included in the power spectrum analysis to $M_{\rm high} = 10^9 M_\odot$ removes an average of $\sim 10$ subhalos. Both the power and the variance are reduced further, but the decrease in variance from changing the highest included mass from $M_{\rm high} = 10^{10} M_\odot$ to $M_{\rm high} = 10^{9} M_\odot$ is not as dramatic as introducing an upper mass limit in the first place. Removing the 1--2 most massive subhalos reduces the variance more than removing the next $\sim 10$.

Making an even more restrictive cut at $M_{\rm high} = 10^8 M_\odot$ removes an average of $\sim 70$ subhalos. While the power is again reduced, the variance of the $M_{\rm high} = 10^8$ and $M_{\rm high} = 10^9$ power spectrum distributions are similar. It appears that the statistical scatter in the power spectrum stabilizes at a mass scale of $10^8$--$10^9 M_\odot$,

\subsection{Projections vs.\ Independent Maps}

To this point we have used each independent population in only one projection while building up the power spectrum distribution. When generating populations is computationally expensive, as with numerical simulations, many projections have been used to estimate the statistical variations \citep[e.g.,][]{ponos_proj}. We can use our set of subhalo populations to test the reliability of using multiple projections of a single population as a proxy for having many independent populations.

From this point onward, we focus our analysis on the central $60 \times 60$ kpc$^2$ of the convergence maps (indicated by the blue squares in \reffig{kapmaps}). Using $500 \times 500$ pixel maps in these regions allows us to reach wavenumbers in the range 0.1--10 kpc$^{-1}$ that can be probed using strong lensing measurements of the power \citep{Pksub}.

\begin{figure*}
\centering
\includegraphics[width=\textwidth]{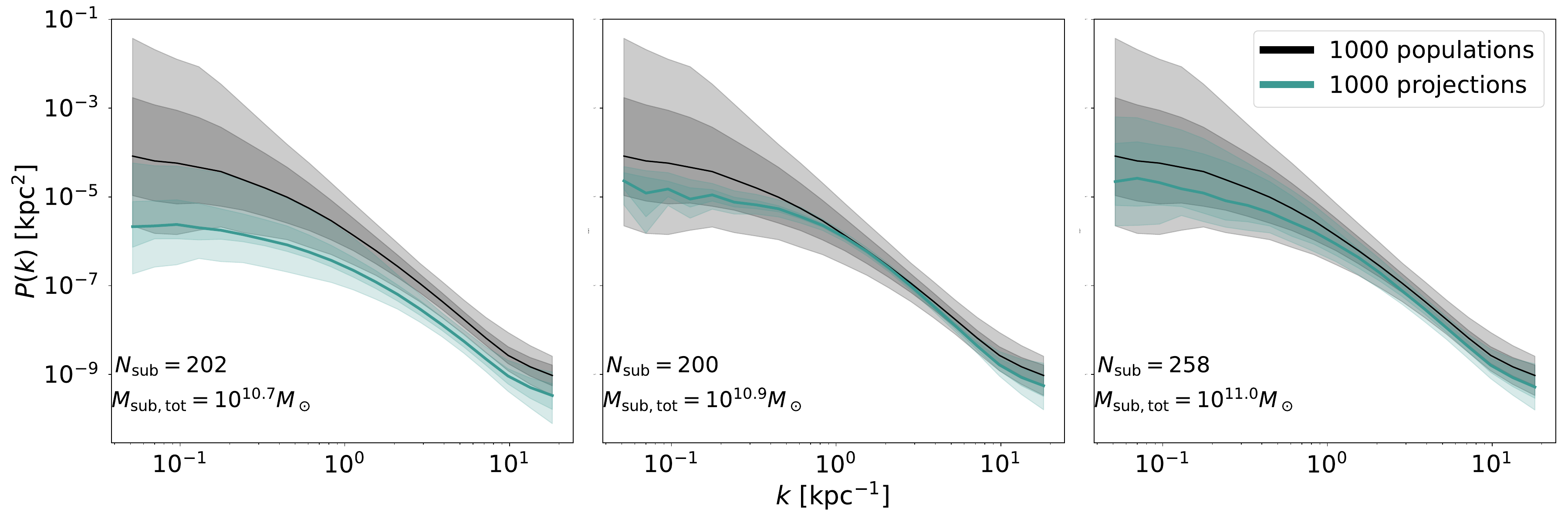}
\caption{Power spectrum distributions of projections vs.\ independent populations. In each panel the solid lines are the median values and the shaded areas are 68 \% and 95\% confidence bands. The gray curves are from a random set of 1000 independent populations and the green curves are from one population projected 1000 times. The flattening at $k \gtrsim 10$ kpc$^{-1}$ is due to reaching the pixel scale of the convergence maps and is not physical. The $N_{\rm sub}$ and $M_{\rm sub, tot}$ values in each panel are the number of subhalos and total mass in subhalos in the projected population shown in that panel.}
\label{proj}
\end{figure*}

\begin{figure*}
\centering
\includegraphics[width=\textwidth]{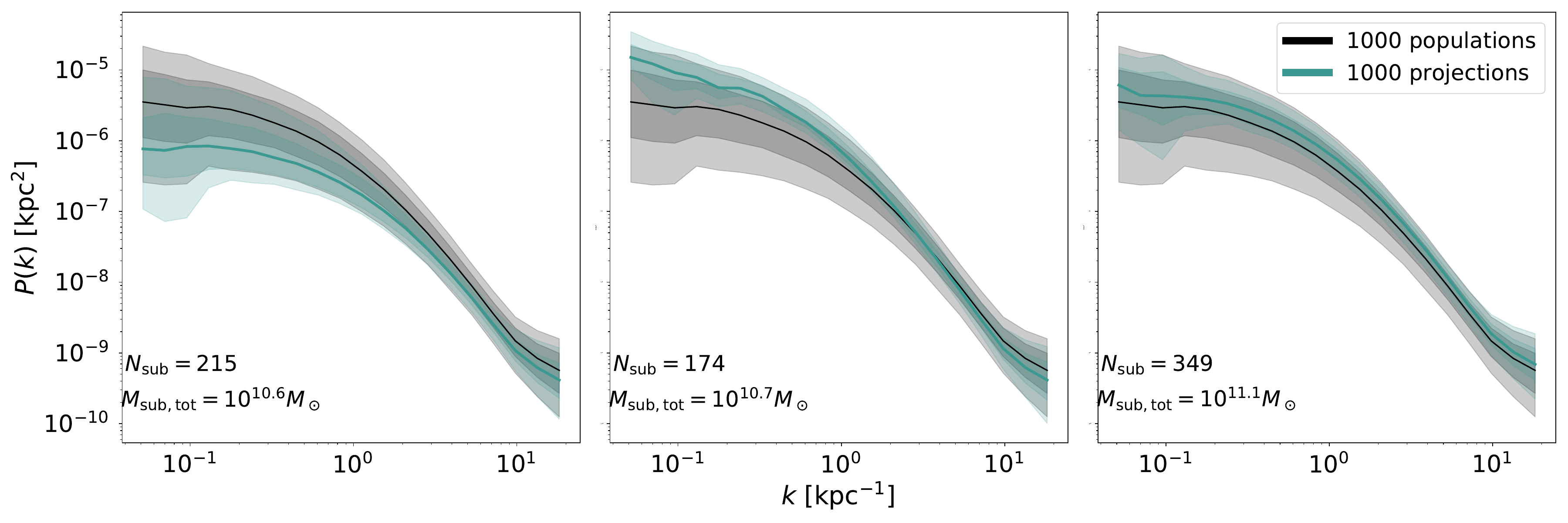}
\caption{Similar to \reffig{proj}, but for populations with subhalos more massive than $10^9 M_\odot$ removed.}
\label{proj_cut}
\end{figure*}

In \reffig{proj} we compare the power spectrum distribution from 1000 independent maps with the distribution from projecting three individual maps 1000 times each (without a mass cut). Projecting a single map multiple times underestimates the variance compared to having an equal number of independent maps. The differences can be understood in terms of effects from massive subhalos. If a population lacks massive subhalos, the power will be low for all projections (as in the left panel of \reffig{proj}). If there is a massive subhalo near the center of the halo, it will appear in the small, central map for most projections, leading to a high power with low variance (as in the middle panel). If there is a massive subhalo at some modest distance from the center, it will sometimes be projected inside the central box within which we compute the power and other times be projected outside the box, leading to a larger variance in the power (as in the right panel of the figure). We note that the assumption of spherical symmetry leads to a reduction in the variance of the power spectrum distribution for different viewing angles compared with a triaxial mass distribution.

The difference between multiple projections and multiple populations is less dramatic when we remove the most massive subhalos, as shown in \reffig{proj_cut}. An upper mass limit of $M_{\rm high} = 10^9 M_\odot$ reduces the scatter among different populations. While the scatter from multiple projections is still somewhat smaller, it is closer to the scatter from multiple populations. We conclude that, apart from the rare massive subhalos, the statistical properties of independent subhalo populations can be approximated by examining many projections of a few populations.

\section{Comparing CDM and WDM}
\label{sec:Results-WDM}

\begin{figure*}
\includegraphics[width=\textwidth]{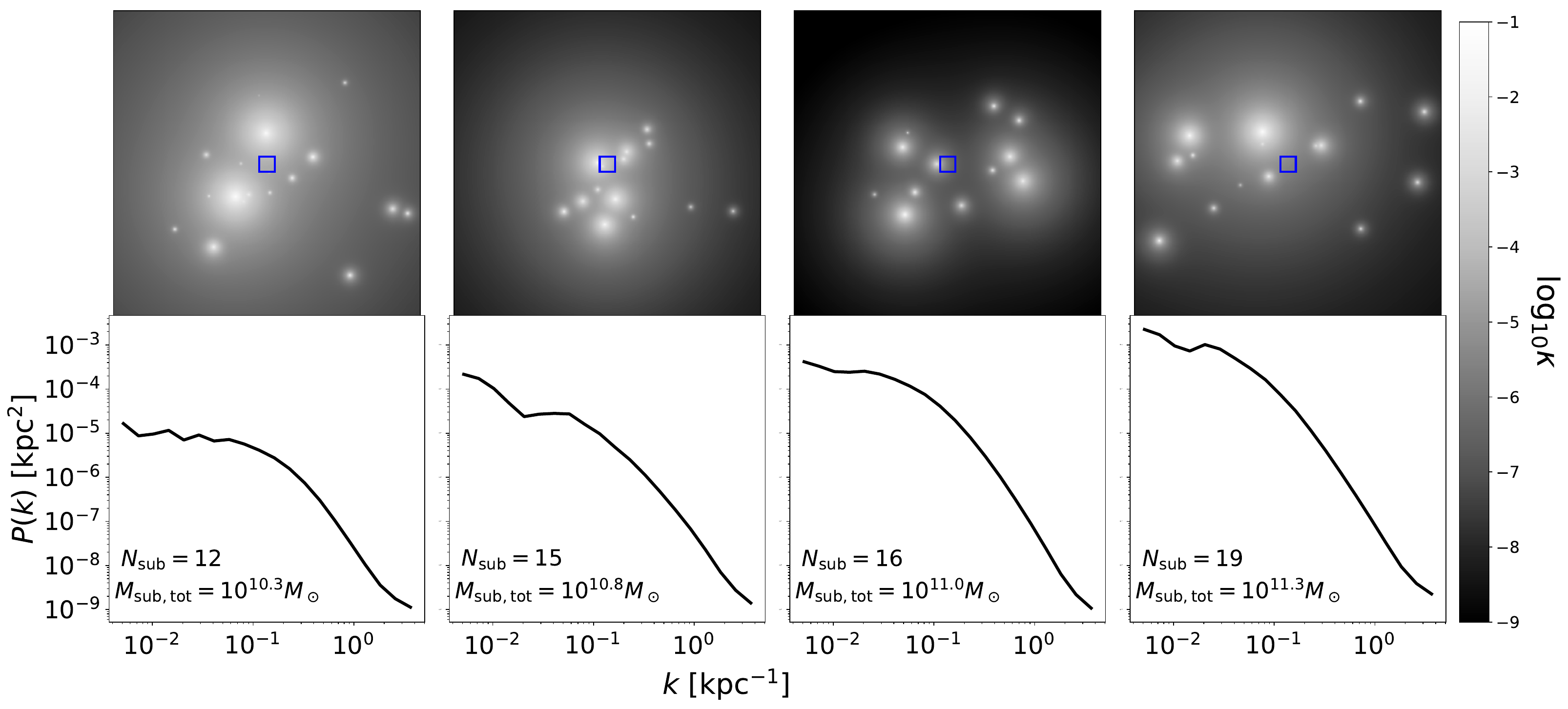}
\caption{
\emph{Top:} Similar to \reffig{kapmaps} for WDM populations assuming a particle mass of 1.5 keV.
\emph{Bottom:} 1-d power spectra corresponding to the maps in the top panels.
}\label{kapmaps_wdm}
\end{figure*}

We are now ready to compare CDM and WDM scenarios using the power spectrum language. Individual WDM maps and their corresponding power spectra are shown in \reffig{kapmaps_wdm}. As a reminder, the WDM particle mass is 1.5 keV. Comparing Figures \ref{kapmaps} with \ref{kapmaps_wdm}, it is immediately apparent that WDM leads to a reduction in the abundance of subhalos, particularly at the low-mass end.

\begin{figure}
\centering
\includegraphics[width=\linewidth]{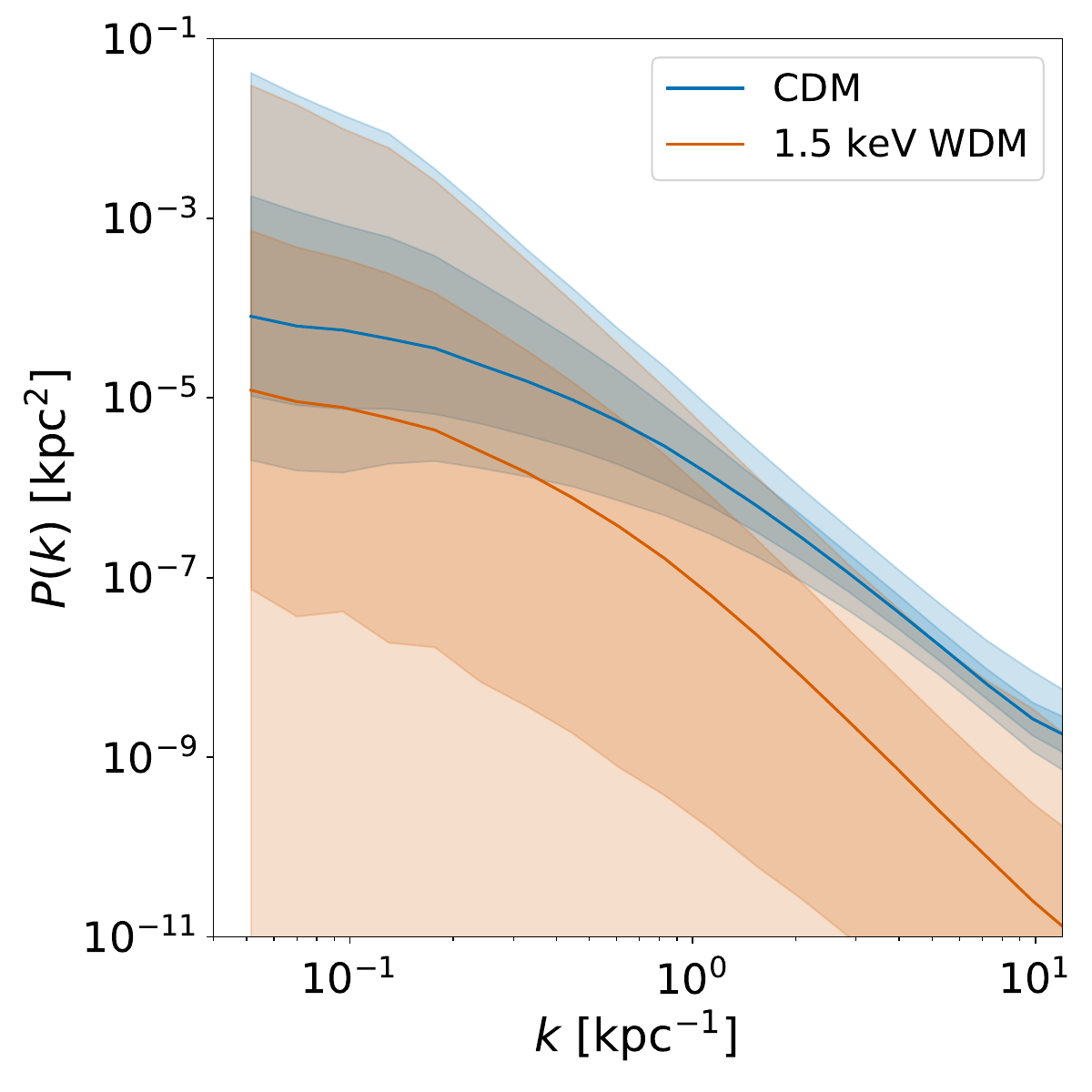}
\caption{Convergence power spectrum distributions for CDM (blue) and 1.5 keV WDM (orange).}
\label{pspec_cdm_wdm}
\end{figure}

In \reffig{pspec_cdm_wdm} we compare the CDM and WDM power spectrum distributions now computed for the small boxes shown by the blue squares in \reffig{kapmaps_wdm}, in order to focus on the range of $k$-values relevant for strong lensing measurements of power). There is considerable overlap between the distributions, and the median values of power are quite similar (especially at $k$-values smaller than shown in the figure). Recall that the amplitude scales roughly as  $P \propto \langle \kappa_{\rm sub} \rangle \langle M^2 \rangle/(\langle M \rangle \Sigma_{\rm crit})$. Looking at \reffig{N_M}, we see that WDM has fewer subhalos but a similar total mass, indicating that the average subhalo mass is higher. In the expression for total power, $\langle \kappa_{\rm sub}\rangle$ is decreased but $\langle M^2 \rangle/\langle M \rangle$ is increased (relative to CDM), leading to a similar overall amplitude for the power spectrum at low $k$. We also note that the WDM power spectrum distribution has a larger scatter that extends to lower power, which is due to Poisson fluctuations in the small number of subhalos in the WDM populations. 

Since WDM suppresses the formation of low-mass structures, we expect the differences between CDM and WDM populations to be most apparent at large $k$-values. At these wavenumbers, where data is likely to display greater sensitivity to the substructure power spectrum (see, e.g., Figure 1 of \citealt{Pksub}), CDM indeed displays significantly more power than WDM in \reffig{pspec_cdm_wdm}, even after accounting for the halo-to-halo scatter. The steeper slope around $k\simeq 1$ kpc$^{-1}$ in the WDM case might provide a way to distinguish it (and other related models displaying a suppressed abundance of small-scale structure) from the standard CDM case. 

\begin{figure}
\includegraphics[width=\linewidth]{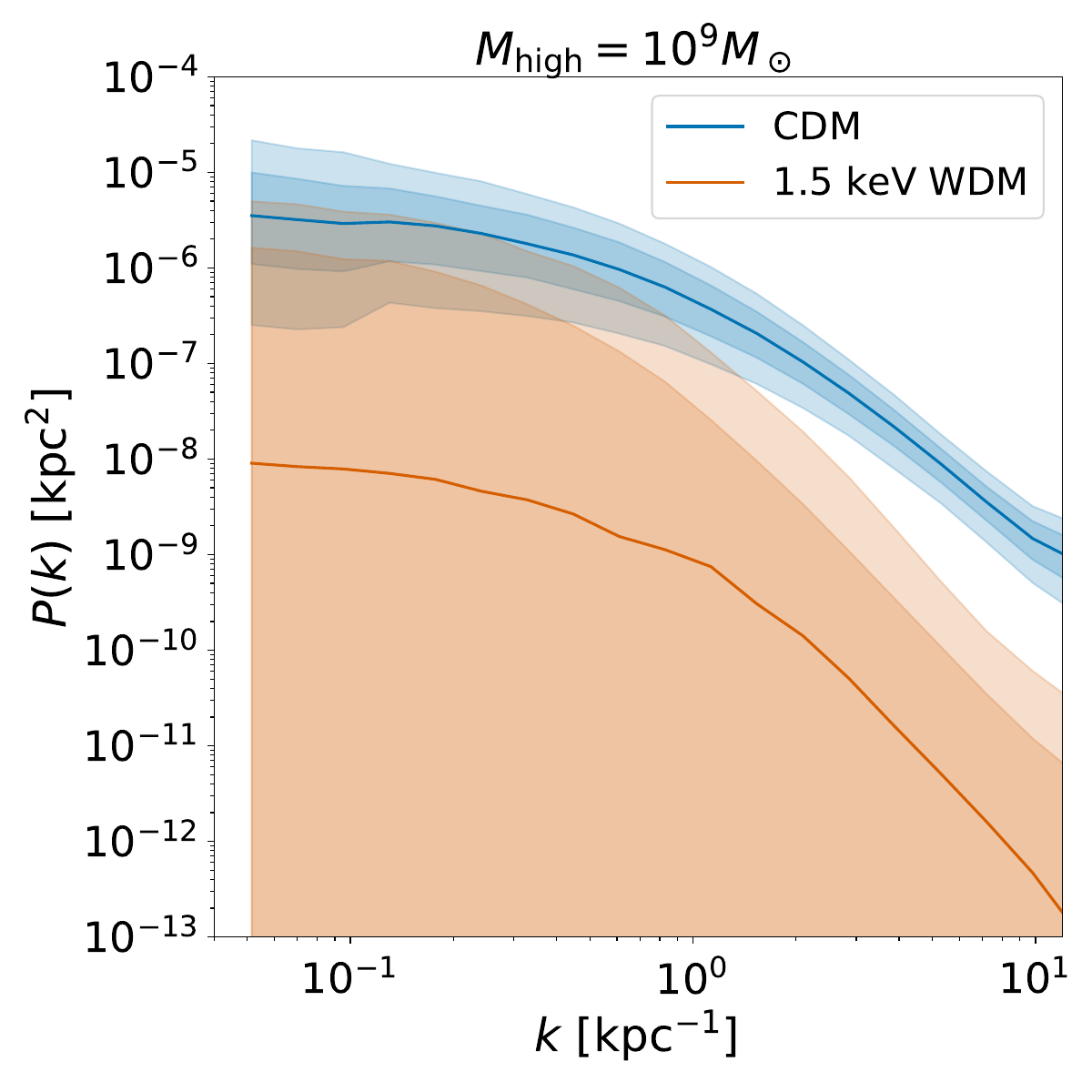}
\caption{Same as \reffig{pspec_cdm_wdm} for populations with halos above $10^9 M_\odot$ removed.}
\label{both_cut}
\end{figure}

Differences between CDM and WDM can be clarified by excluding the most massive subhalos from the power spectrum calculation (and treating them explicitly in the lens mass model instead). In \reffig{both_cut} we compare CDM and WDM power spectrum distributions for populations with a highest allowed mass of $M_{\rm high} = 10^9 M_\odot$. We see that the power spectra look quite different on all scales in this case. In WDM, including only subhalos with mass below $10^9 M_\odot$ removes a higher fraction of the total number of subhalos, and thus more of the total power, compared to CDM. The differences in amplitude and slope at $k\simeq 1$ kpc$^{-1}$ are again the most striking features of these power spectra. The wavenumber range 0.1--2 kpc$^{-1}$ relevant for strong lensing probes subhalo masses in the $\sim 10^9$--$10^{10} M_{\odot}$ range, which provides sensitivity to WDM particle masses of $\sim$1--3 keV. Our results therefore indicate that strong lensing measurements of small-scale power could constrain WDM particle masses in the range of a few keV \citep[also see][]{gilman,gilman2019}.

\section{Conclusions}
\label{sec:Conclusions}

We have computed the convergence power spectrum of dark matter substructure using semi-analytic subhalo populations in both cold and warm dark matter scenarios. The power spectrum distributions for CDM and WDM have similar shapes and overall levels of power at low wavenumbers, but the scatter appears larger for WDM. The scatter in the power spectrum distribution is driven by the few most massive subhalos. Those subhalos could potentially be individually detected and directly included in the main lens model, so they can be excluded from the power spectrum analysis. When that is done, the resulting power spectrum distributions are statistically robust and show clear differences between CDM and WDM predictions on scales $k \gtrsim 0.1$ kpc$^{-1}$.

This result is promising in connection with recent work on using galaxy-scale strong lensing to measure small-scale power. \citet{Pksub} recently developed a comprehensive likelihood-based formalism and used it to demonstrate that measuring power on scales of $k \sim 0.1$--10 kpc$^{-1}$ with lensing is likely feasible with deep, high-resolution observations. Our analysis indicates that even a few high-quality power spectrum measurements in this $k$ range could be sufficient to measure potential deviations from the CDM predictions for dark matter substructure within galaxies.

The simulations used here contain only dark matter, but a similar analysis could be used to study the effects of baryons on the small-scale power spectrum. While full $N$-body simulations with baryons are computationally challenging, semi-analytic modeling offers a useful alternative due to the relative ease of including baryonic structures like disks and bulges. Such structures would likely contribute power at intermediate scales where CDM and WDM power spectra begin to differ, so it will be important to understand how baryons affect the power spectrum in the $k \sim 0.1$--$10$ kpc$^{-1}$ range if we want to use strong lensing measurements of the power spectrum to distinguish dark matter models.

\section*{Acknowledgements}

S.B.\ and C.R.K.\ acknowledge the support of grant AST-1716585 from the National Science Foundation, as well as programs HST-AR-14305.002-A and HST-AR-15007.002-A whose support was provided by the National Aeronautical and Space Administration (NASA) through a grant from the Space Telescope Science Institute, which is operated by the Association of Universities for Research in Astronomy, Incorporated, under NASA contract NAS5-26555. F.-Y. C.-R.~acknowledges the support of the NASA ATP grant NNX16AI12G at Harvard University. Part of this work took place at the Jet Propulsion Laboratory, California Institute of Technology, under a contract with NASA.

\bibliographystyle{mnras}
\bibliography{bib}

\begin{thebibliography}{}
\makeatletter
\relax
\def\mn@urlcharsother{\let\do\@makeother \do\$\do\&\do\#\do\^\do\_\do\%\do\~}
\def\mn@doi{\begingroup\mn@urlcharsother \@ifnextchar [ {\mn@doi@}
  {\mn@doi@[]}}
\def\mn@doi@[#1]#2{\def\@tempa{#1}\ifx\@tempa\@empty \href
  {http://dx.doi.org/#2} {doi:#2}\else \href {http://dx.doi.org/#2} {#1}\fi
  \endgroup}
\def\mn@eprint#1#2{\mn@eprint@#1:#2::\@nil}
\def\mn@eprint@arXiv#1{\href {http://arxiv.org/abs/#1} {{\tt arXiv:#1}}}
\def\mn@eprint@dblp#1{\href {http://dblp.uni-trier.de/rec/bibtex/#1.xml}
  {dblp:#1}}
\def\mn@eprint@#1:#2:#3:#4\@nil{\def\@tempa {#1}\def\@tempb {#2}\def\@tempc
  {#3}\ifx \@tempc \@empty \let \@tempc \@tempb \let \@tempb \@tempa \fi \ifx
  \@tempb \@empty \def\@tempb {arXiv}\fi \@ifundefined
  {mn@eprint@\@tempb}{\@tempb:\@tempc}{\expandafter \expandafter \csname
  mn@eprint@\@tempb\endcsname \expandafter{\@tempc}}}

\bibitem[\protect\citeauthoryear{Anderhalden, Schneider, Maccio, Diemand  \&
  Bertone}{Anderhalden et~al.}{2013}]{Anderhalden:2012jc}
Anderhalden D.,  Schneider A.,  Maccio A.~V.,  Diemand J.,   Bertone G.,  2013,
  \mn@doi [JCAP] {10.1088/1475-7516/2013/03/014}, 1303, 014

\bibitem[\protect\citeauthoryear{{Baltz}, {Marshall}  \& {Oguri}}{{Baltz}
  et~al.}{2009}]{baltz}
{Baltz} E.~A.,  {Marshall} P.,   {Oguri} M.,  2009, \mn@doi [JCAP]
  {10.1088/1475-7516/2009/01/015}, \href
  {http://adsabs.harvard.edu/abs/2009JCAP...01..015B} {1, 015}

\bibitem[\protect\citeauthoryear{{Banik}, {Bertone}, {Bovy}  \&
  {Bozorgnia}}{{Banik} et~al.}{2018}]{Banik:2018pjp}
{Banik} N.,  {Bertone} G.,  {Bovy} J.,   {Bozorgnia} N.,  2018, \mn@doi [\jcap]
  {10.1088/1475-7516/2018/07/061}, \href
  {http://adsabs.harvard.edu/abs/2018JCAP...07..061B} {7, 061}

\bibitem[\protect\citeauthoryear{{Bayer}, {Chatterjee}, {Koopmans}, {Vegetti},
  {McKean}, {Treu}  \& {Fassnacht}}{{Bayer} et~al.}{2018}]{Bayer}
{Bayer} D.,  {Chatterjee} S.,  {Koopmans} L.~V.~E.,  {Vegetti} S.,  {McKean}
  J.~P.,  {Treu} T.,   {Fassnacht} C.~D.,  2018, preprint, \href
  {http://adsabs.harvard.edu/abs/2018arXiv180305952B} {} (\mn@eprint {arXiv}
  {1803.05952})

\bibitem[\protect\citeauthoryear{Behroozi, Wechsler  \& Conroy}{Behroozi
  et~al.}{2013}]{Behroozi:2012iw}
Behroozi P.~S.,  Wechsler R.~H.,   Conroy C.,  2013, \mn@doi [\apj]
  {10.1088/0004-637X/770/1/57}, 770, 57

\bibitem[\protect\citeauthoryear{{Benson}}{{Benson}}{2012}]{galacticus}
{Benson} A.~J.,  2012, \mn@doi [NewA] {10.1016/j.newast.2011.07.004}, \href
  {http://adsabs.harvard.edu/abs/2012NewA...17..175B} {17, 175}

\bibitem[\protect\citeauthoryear{{Benson}, {Lacey}, {Baugh}, {Cole}  \&
  {Frenk}}{{Benson} et~al.}{2002}]{Benson02}
{Benson} A.~J.,  {Lacey} C.~G.,  {Baugh} C.~M.,  {Cole} S.,   {Frenk} C.~S.,
  2002, \mn@doi [\mnras] {10.1046/j.1365-8711.2002.05387.x}, \href
  {http://adsabs.harvard.edu/abs/2002MNRAS.333..156B} {333, 156}

\bibitem[\protect\citeauthoryear{{Birrer}, {Amara}  \& {Refregier}}{{Birrer}
  et~al.}{2017}]{Birrer2017}
{Birrer} S.,  {Amara} A.,   {Refregier} A.,  2017, \mn@doi [JCAP]
  {10.1088/1475-7516/2017/05/037}, \href
  {http://adsabs.harvard.edu/abs/2017JCAP...05..037B} {5, 037}

\bibitem[\protect\citeauthoryear{Bose et~al.,}{Bose
  et~al.}{2017}]{Bose:2016irl}
Bose S.,  et~al., 2017, \mn@doi [MNRAS] {10.1093/mnras/stw2686}, 464, 4520

\bibitem[\protect\citeauthoryear{{Bovy}}{{Bovy}}{2016}]{2016PhRvL.116l1301B}
{Bovy} J.,  2016, \mn@doi [Physical Review Letters]
  {10.1103/PhysRevLett.116.121301}, \href
  {http://adsabs.harvard.edu/abs/2016PhRvL.116l1301B} {116, 121301}

\bibitem[\protect\citeauthoryear{{Bovy}, {Erkal}  \& {Sanders}}{{Bovy}
  et~al.}{2017}]{2017MNRAS.466..628B}
{Bovy} J.,  {Erkal} D.,   {Sanders} J.~L.,  2017, \mn@doi [\mnras]
  {10.1093/mnras/stw3067}, \href
  {http://adsabs.harvard.edu/abs/2017MNRAS.466..628B} {466, 628}

\bibitem[\protect\citeauthoryear{{Boylan-Kolchin}, {Springel}, {White},
  {Jenkins}  \& {Lemson}}{{Boylan-Kolchin} et~al.}{2009}]{boylan}
{Boylan-Kolchin} M.,  {Springel} V.,  {White} S.~D.~M.,  {Jenkins} A.,
  {Lemson} G.,  2009, \mn@doi [MNRAS] {10.1111/j.1365-2966.2009.15191.x}, \href
  {http://adsabs.harvard.edu/abs/2009MNRAS.398.1150B} {398, 1150}

\bibitem[\protect\citeauthoryear{Brook, Di~Cintio, Knebe, Gottlöber, Hoffman,
  Yepes  \& Garrison-Kimmel}{Brook et~al.}{2014}]{Brook:2013laa}
Brook C.~B.,  Di~Cintio A.,  Knebe A.,  Gottlöber S.,  Hoffman Y.,  Yepes G.,
   Garrison-Kimmel S.,  2014, \mn@doi [\apj] {10.1088/2041-8205/784/1/L14},
  784, L14

\bibitem[\protect\citeauthoryear{{Brooks}, {Kuhlen}, {Zolotov}  \&
  {Hooper}}{{Brooks} et~al.}{2013}]{Brooks}
{Brooks} A.~M.,  {Kuhlen} M.,  {Zolotov} A.,   {Hooper} D.,  2013, \mn@doi
  [\apj] {10.1088/0004-637X/765/1/22}, \href
  {http://adsabs.harvard.edu/abs/2013ApJ...765...22B} {765, 22}

\bibitem[\protect\citeauthoryear{{Bullock}, {Kravtsov}  \&
  {Weinberg}}{{Bullock} et~al.}{2000}]{Bullock}
{Bullock} J.~S.,  {Kravtsov} A.~V.,   {Weinberg} D.~H.,  2000, \mn@doi [\apj]
  {10.1086/309279}, \href {http://adsabs.harvard.edu/abs/2000ApJ...539..517B}
  {539, 517}

\bibitem[\protect\citeauthoryear{{Buschmann}, {Kopp}, {Safdi}  \&
  {Wu}}{{Buschmann} et~al.}{2018}]{Buschmann:2017ams}
{Buschmann} M.,  {Kopp} J.,  {Safdi} B.~R.,   {Wu} C.-L.,  2018, \mn@doi
  [Physical Review Letters] {10.1103/PhysRevLett.120.211101}, \href
  {http://adsabs.harvard.edu/abs/2018PhRvL.120u1101B} {120, 211101}

\bibitem[\protect\citeauthoryear{{Carlberg}}{{Carlberg}}{2016}]{2016ApJ...820...45C}
{Carlberg} R.~G.,  2016, \mn@doi [\apj] {10.3847/0004-637X/820/1/45}, \href
  {http://adsabs.harvard.edu/abs/2016ApJ...820...45C} {820, 45}

\bibitem[\protect\citeauthoryear{{Chatterjee} \& {Koopmans}}{{Chatterjee} \&
  {Koopmans}}{2018}]{Chatterjee}
{Chatterjee} S.,  {Koopmans} L.~V.~E.,  2018, \mn@doi [\mnras]
  {10.1093/mnras/stx2674}, \href
  {http://adsabs.harvard.edu/abs/2018MNRAS.474.1762C} {474, 1762}

\bibitem[\protect\citeauthoryear{{Cyr-Racine}, {Moustakas}, {Keeton},
  {Sigurdson}  \& {Gilman}}{{Cyr-Racine} et~al.}{2016}]{FYCR}
{Cyr-Racine} F.-Y.,  {Moustakas} L.~A.,  {Keeton} C.~R.,  {Sigurdson} K.,
  {Gilman} D.~A.,  2016, \mn@doi [Phys. Rev. D] {10.1103/PhysRevD.94.043505},
  \href {http://adsabs.harvard.edu/abs/2016PhRvD..94d3505C} {94, 043505}

\bibitem[\protect\citeauthoryear{{Cyr-Racine}, {Keeton}  \&
  {Moustakas}}{{Cyr-Racine} et~al.}{2018}]{Pksub}
{Cyr-Racine} F.-Y.,  {Keeton} C.~R.,   {Moustakas} L.~A.,  2018, preprint,
  \href {http://adsabs.harvard.edu/abs/2018arXiv180607897C} {} (\mn@eprint
  {arXiv} {1806.07897})

\bibitem[\protect\citeauthoryear{{Dalal} \& {Kochanek}}{{Dalal} \&
  {Kochanek}}{2002}]{dalal_kochanek}
{Dalal} N.,  {Kochanek} C.~S.,  2002, \mn@doi [\apj] {10.1086/340303}, \href
  {http://adsabs.harvard.edu/abs/2002ApJ...572...25D} {572, 25}

\bibitem[\protect\citeauthoryear{Daylan, Cyr-Racine, Diaz~Rivero, Dvorkin  \&
  Finkbeiner}{Daylan et~al.}{2018}]{Daylan:2017kfh}
Daylan T.,  Cyr-Racine F.-Y.,  Diaz~Rivero A.,  Dvorkin C.,   Finkbeiner D.~P.,
   2018, \mn@doi [\apj] {10.3847/1538-4357/aaaa1e}, 854, 141

\bibitem[\protect\citeauthoryear{Despali, Vegetti, White, Giocoli  \& van~den
  Bosch}{Despali et~al.}{2018}]{Despali:2017ksx}
Despali G.,  Vegetti S.,  White S. D.~M.,  Giocoli C.,   van~den Bosch F.~C.,
  2018, \mn@doi [MNRAS] {10.1093/mnras/sty159}, 475, 5424

\bibitem[\protect\citeauthoryear{{Diaz Rivero}, {Cyr-Racine}  \&
  {Dvorkin}}{{Diaz Rivero} et~al.}{2017}]{Diaz}
{Diaz Rivero} A.,  {Cyr-Racine} F.-Y.,   {Dvorkin} C.,  2017, preprint, \href
  {http://adsabs.harvard.edu/abs/2017arXiv170704590D} {} (\mn@eprint {arXiv}
  {1707.04590})

\bibitem[\protect\citeauthoryear{Dooley, Peter, Yang, Willman, Griffen  \&
  Frebel}{Dooley et~al.}{2017}]{Dooley:2016xkj}
Dooley G.~A.,  Peter A. H.~G.,  Yang T.,  Willman B.,  Griffen B.~F.,   Frebel
  A.,  2017, \mn@doi [\mnras] {10.1093/mnras/stx1900}, 471, 4894

\bibitem[\protect\citeauthoryear{Erickcek \& Law}{Erickcek \&
  Law}{2011}]{Erickcek:2010fc}
Erickcek A.~L.,  Law N.~M.,  2011, \mn@doi [\apj] {10.1088/0004-637X/729/1/49},
  729, 49

\bibitem[\protect\citeauthoryear{{Erkal}, {Belokurov}, {Bovy}  \&
  {Sanders}}{{Erkal} et~al.}{2016}]{2016MNRAS.463..102E}
{Erkal} D.,  {Belokurov} V.,  {Bovy} J.,   {Sanders} J.~L.,  2016, \mn@doi
  [\mnras] {10.1093/mnras/stw1957}, \href
  {http://adsabs.harvard.edu/abs/2016MNRAS.463..102E} {463, 102}

\bibitem[\protect\citeauthoryear{Escudero, Lopez-Honorez, Mena, Palomares-Ruiz
  \& Villanueva-Domingo}{Escudero et~al.}{2018}]{Escudero:2018thh}
Escudero M.,  Lopez-Honorez L.,  Mena O.,  Palomares-Ruiz S.,
  Villanueva-Domingo P.,  2018, \mn@doi [JCAP] {10.1088/1475-7516/2018/06/007},
  1806, 007

\bibitem[\protect\citeauthoryear{{Fadely} \& {Keeton}}{{Fadely} \&
  {Keeton}}{2012}]{fadely}
{Fadely} R.,  {Keeton} C.~R.,  2012, \mn@doi [\mnras]
  {10.1111/j.1365-2966.2011.19729.x}, \href
  {http://adsabs.harvard.edu/abs/2012MNRAS.419..936F} {419, 936}

\bibitem[\protect\citeauthoryear{Feldmann \& Spolyar}{Feldmann \&
  Spolyar}{2015}]{Feldmann:2013hqa}
Feldmann R.,  Spolyar D.,  2015, \mn@doi [\mnras] {10.1093/mnras/stu2147}, 446,
  1000

\bibitem[\protect\citeauthoryear{{Fiacconi}, {Madau}, {Potter}  \&
  {Stadel}}{{Fiacconi} et~al.}{2016}]{ponos_proj}
{Fiacconi} D.,  {Madau} P.,  {Potter} D.,   {Stadel} J.,  2016, \mn@doi [\apj]
  {10.3847/0004-637X/824/2/144}, \href
  {http://adsabs.harvard.edu/abs/2016ApJ...824..144F} {824, 144}

\bibitem[\protect\citeauthoryear{{Gao}, {Frenk}, {Boylan-Kolchin}, {Jenkins},
  {Springel}  \& {White}}{{Gao} et~al.}{2011}]{z_evol}
{Gao} L.,  {Frenk} C.~S.,  {Boylan-Kolchin} M.,  {Jenkins} A.,  {Springel} V.,
   {White} S.~D.~M.,  2011, \mn@doi [\mnras]
  {10.1111/j.1365-2966.2010.17601.x}, \href
  {http://adsabs.harvard.edu/abs/2011MNRAS.410.2309G} {410, 2309}

\bibitem[\protect\citeauthoryear{Garrison-Kimmel, Boylan-Kolchin, Bullock  \&
  Lee}{Garrison-Kimmel et~al.}{2014}]{Garrison-Kimmel:2013eoa}
Garrison-Kimmel S.,  Boylan-Kolchin M.,  Bullock J.,   Lee K.,  2014, \mn@doi
  [\mnras] {10.1093/mnras/stt2377}, 438, 2578

\bibitem[\protect\citeauthoryear{{Gilman}, {Birrer}, {Treu}, {Keeton}  \&
  {Nierenberg}}{{Gilman} et~al.}{2018}]{gilman}
{Gilman} D.,  {Birrer} S.,  {Treu} T.,  {Keeton} C.~R.,   {Nierenberg} A.,
  2018, \mn@doi [\mnras] {10.1093/mnras/sty2261}, \href
  {http://adsabs.harvard.edu/abs/2018MNRAS.481..819G} {481, 819}

\bibitem[\protect\citeauthoryear{{Gilman}, {Birrer}, {Treu}, {Nierenberg}  \&
  {Benson}}{{Gilman} et~al.}{2019}]{gilman2019}
{Gilman} D.,  {Birrer} S.,  {Treu} T.,  {Nierenberg} A.,   {Benson} A.,  2019,
  arXiv e-prints, \href {http://adsabs.harvard.edu/abs/2019arXiv190111031G} {}

\bibitem[\protect\citeauthoryear{{G{\"o}tz} \& {Sommer-Larsen}}{{G{\"o}tz} \&
  {Sommer-Larsen}}{2002}]{wdm_goetz}
{G{\"o}tz} M.,  {Sommer-Larsen} J.,  2002, \mn@doi [\apss]
  {10.1023/A:1019543230202}, \href
  {http://adsabs.harvard.edu/abs/2002Ap%26SS.281..415G} {281, 415}

\bibitem[\protect\citeauthoryear{{Governato} et~al.,}{{Governato}
  et~al.}{2015}]{2015MNRAS.448..792G}
{Governato} F.,  et~al., 2015, \mn@doi [\mnras] {10.1093/mnras/stu2720}, \href
  {http://adsabs.harvard.edu/abs/2015MNRAS.448..792G} {448, 792}

\bibitem[\protect\citeauthoryear{Hezaveh, Dalal, Holder, Kuhlen, Marrone
  et~al.}{Hezaveh et~al.}{2013}]{Hezaveh:2012ai}
Hezaveh Y.,  Dalal N.,  Holder G.,  Kuhlen M.,  Marrone D.,   et~al., 2013,
  \mn@doi [ApJ] {10.1088/0004-637X/767/1/9}, 767, 9

\bibitem[\protect\citeauthoryear{{Hezaveh}, {Dalal}, {Holder}, {Kisner},
  {Kuhlen}  \& {Perreault Levasseur}}{{Hezaveh} et~al.}{2016a}]{hez_pow}
{Hezaveh} Y.,  {Dalal} N.,  {Holder} G.,  {Kisner} T.,  {Kuhlen} M.,
  {Perreault Levasseur} L.,  2016a, \mn@doi [JCAP]
  {10.1088/1475-7516/2016/11/048}, \href
  {http://adsabs.harvard.edu/abs/2016JCAP...11..048H} {11, 048}

\bibitem[\protect\citeauthoryear{{Hezaveh} et~al.,}{{Hezaveh}
  et~al.}{2016b}]{hez_clump}
{Hezaveh} Y.~D.,  et~al., 2016b, \mn@doi [ApJ] {10.3847/0004-637X/823/1/37},
  \href {http://adsabs.harvard.edu/abs/2016ApJ...823...37H} {823, 37}

\bibitem[\protect\citeauthoryear{{Hsueh}, {Fassnacht}, {Vegetti}, {McKean},
  {Spingola}, {Auger}, {Koopmans}  \& {Lagattuta}}{{Hsueh}
  et~al.}{2016}]{Hsueh_16}
{Hsueh} J.-W.,  {Fassnacht} C.~D.,  {Vegetti} S.,  {McKean} J.~P.,  {Spingola}
  C.,  {Auger} M.~W.,  {Koopmans} L.~V.~E.,   {Lagattuta} D.~J.,  2016, \mn@doi
  [\mnras] {10.1093/mnrasl/slw146}, \href
  {http://adsabs.harvard.edu/abs/2016MNRAS.463L..51H} {463, L51}

\bibitem[\protect\citeauthoryear{{Hsueh} et~al.,}{{Hsueh}
  et~al.}{2017}]{Hsueh_17}
{Hsueh} J.-W.,  et~al., 2017, \mn@doi [\mnras] {10.1093/mnras/stx1082}, \href
  {http://adsabs.harvard.edu/abs/2017MNRAS.469.3713H} {469, 3713}

\bibitem[\protect\citeauthoryear{{Hsueh}, {Despali}, {Vegetti}, {Xu},
  {Fassnacht}  \& {Metcalf}}{{Hsueh} et~al.}{2018}]{Hsueh_18}
{Hsueh} J.-W.,  {Despali} G.,  {Vegetti} S.,  {Xu} D.,  {Fassnacht} C.~D.,
  {Metcalf} R.~B.,  2018, \mn@doi [\mnras] {10.1093/mnras/stx3320}, \href
  {http://adsabs.harvard.edu/abs/2018MNRAS.475.2438H} {475, 2438}

\bibitem[\protect\citeauthoryear{Iršič et~al.}{Iršič
  et~al.}{2017}]{Irsic:2017ixq}
Iršič V.,  et~al., 2017, \mn@doi [Phys. Rev.] {10.1103/PhysRevD.96.023522},
  D96, 023522

\bibitem[\protect\citeauthoryear{Keeton}{Keeton}{2003}]{Keeton:2003aa}
Keeton C.~R.,  2003, Astrophys. J., 584, 664

\bibitem[\protect\citeauthoryear{{Keeton}, {Gaudi}  \& {Petters}}{{Keeton}
  et~al.}{2003}]{kgp_cusp}
{Keeton} C.~R.,  {Gaudi} B.~S.,   {Petters} A.~O.,  2003, \mn@doi [\apj]
  {10.1086/378934}, \href {http://adsabs.harvard.edu/abs/2003ApJ...598..138K}
  {598, 138}

\bibitem[\protect\citeauthoryear{{Keeton}, {Gaudi}  \& {Petters}}{{Keeton}
  et~al.}{2005}]{kgp_fold}
{Keeton} C.~R.,  {Gaudi} B.~S.,   {Petters} A.~O.,  2005, \mn@doi [\apj]
  {10.1086/497324}, \href {http://adsabs.harvard.edu/abs/2005ApJ...635...35K}
  {635, 35}

\bibitem[\protect\citeauthoryear{{Kim}, {Peter}  \& {Hargis}}{{Kim}
  et~al.}{2017}]{Kim:2017iwr}
{Kim} S.~Y.,  {Peter} A.~H.~G.,   {Hargis} J.~R.,  2017, arXiv e-prints, \href
  {http://adsabs.harvard.edu/abs/2017arXiv171106267K} {}

\bibitem[\protect\citeauthoryear{{Koopmans}}{{Koopmans}}{2005}]{Koopmans:aa}
{Koopmans} L.~V.~E.,  2005, \mn@doi [\mnras]
  {10.1111/j.1365-2966.2005.09523.x}, \href
  {http://adsabs.harvard.edu/abs/2005MNRAS.363.1136K} {363, 1136}

\bibitem[\protect\citeauthoryear{Lovell, Frenk, Eke, Jenkins, Gao  \&
  Theuns}{Lovell et~al.}{2014}]{Lovell:2013ola}
Lovell M.~R.,  Frenk C.~S.,  Eke V.~R.,  Jenkins A.,  Gao L.,   Theuns T.,
  2014, \mn@doi [MNRAS] {10.1093/mnras/stt2431}, 439, 300

\bibitem[\protect\citeauthoryear{{Mao} \& {Schneider}}{{Mao} \&
  {Schneider}}{1998}]{Mao98}
{Mao} S.,  {Schneider} P.,  1998, \mn@doi [\mnras]
  {10.1046/j.1365-8711.1998.01319.x}, \href
  {http://adsabs.harvard.edu/abs/1998MNRAS.295..587M} {295, 587}

\bibitem[\protect\citeauthoryear{{Metcalf} \& {Madau}}{{Metcalf} \&
  {Madau}}{2001}]{metcalf01}
{Metcalf} R.~B.,  {Madau} P.,  2001, \mn@doi [\apj] {10.1086/323695}, \href
  {http://adsabs.harvard.edu/abs/2001ApJ...563....9M} {563, 9}

\bibitem[\protect\citeauthoryear{Minor, Kaplinghat  \& Li}{Minor
  et~al.}{2017}]{Minor:2016jou}
Minor Q.~E.,  Kaplinghat M.,   Li N.,  2017, \mn@doi [\apj]
  {10.3847/1538-4357/aa7fee}, 845, 118

\bibitem[\protect\citeauthoryear{{Ngan} \& {Carlberg}}{{Ngan} \&
  {Carlberg}}{2014}]{2014ApJ...788..181N}
{Ngan} W.~H.~W.,  {Carlberg} R.~G.,  2014, \mn@doi [\apj]
  {10.1088/0004-637X/788/2/181}, \href
  {http://adsabs.harvard.edu/abs/2014ApJ...788..181N} {788, 181}

\bibitem[\protect\citeauthoryear{{Nierenberg}, {Treu}, {Wright}, {Fassnacht}
  \& {Auger}}{{Nierenberg} et~al.}{2014}]{Nierenberg:2014aa}
{Nierenberg} A.~M.,  {Treu} T.,  {Wright} S.~A.,  {Fassnacht} C.~D.,   {Auger}
  M.~W.,  2014, \mn@doi [\mnras] {10.1093/mnras/stu862}, \href
  {http://adsabs.harvard.edu/abs/2014MNRAS.442.2434N} {442, 2434}

\bibitem[\protect\citeauthoryear{{Pullen}, {Benson}  \& {Moustakas}}{{Pullen}
  et~al.}{2014}]{pullen}
{Pullen} A.~R.,  {Benson} A.~J.,   {Moustakas} L.~A.,  2014, \mn@doi [ApJ]
  {10.1088/0004-637X/792/1/24}, \href
  {http://adsabs.harvard.edu/abs/2014ApJ...792...24P} {792, 24}

\bibitem[\protect\citeauthoryear{{Rodr{\'{\i}}guez-Puebla}, {Primack},
  {Avila-Reese}  \& {Faber}}{{Rodr{\'{\i}}guez-Puebla}
  et~al.}{2017}]{2017MNRAS.470..651R}
{Rodr{\'{\i}}guez-Puebla} A.,  {Primack} J.~R.,  {Avila-Reese} V.,   {Faber}
  S.~M.,  2017, \mn@doi [\mnras] {10.1093/mnras/stx1172}, \href
  {http://adsabs.harvard.edu/abs/2017MNRAS.470..651R} {470, 651}

\bibitem[\protect\citeauthoryear{Schneider}{Schneider}{2015}]{Schneider:2014rda}
Schneider A.,  2015, \mn@doi [\mnras] {10.1093/mnras/stv1169}, 451, 3117

\bibitem[\protect\citeauthoryear{{Somerville}, {Bullock}  \&
  {Livio}}{{Somerville} et~al.}{2003}]{Somerville}
{Somerville} R.~S.,  {Bullock} J.~S.,   {Livio} M.,  2003, \mn@doi [\apj]
  {10.1086/376686}, \href {http://adsabs.harvard.edu/abs/2003ApJ...593..616S}
  {593, 616}

\bibitem[\protect\citeauthoryear{{Springel} et~al.,}{{Springel}
  et~al.}{2008}]{springel}
{Springel} V.,  et~al., 2008, \mn@doi [\mnras]
  {10.1111/j.1365-2966.2008.14066.x}, \href
  {http://adsabs.harvard.edu/abs/2008MNRAS.391.1685S} {391, 1685}

\bibitem[\protect\citeauthoryear{{Van Tilburg}, {Taki}  \& {Weiner}}{{Van
  Tilburg} et~al.}{2018}]{VanTilburg:2018ykj}
{Van Tilburg} K.,  {Taki} A.-M.,   {Weiner} N.,  2018, \mn@doi [\jcap]
  {10.1088/1475-7516/2018/07/041}, \href
  {http://adsabs.harvard.edu/abs/2018JCAP...07..041V} {7, 041}

\bibitem[\protect\citeauthoryear{{Vegetti} \& {Koopmans}}{{Vegetti} \&
  {Koopmans}}{2009}]{Vegetti:2008aa}
{Vegetti} S.,  {Koopmans} L.~V.~E.,  2009, \mn@doi [\mnras]
  {10.1111/j.1365-2966.2008.14005.x}, \href
  {http://adsabs.harvard.edu/abs/2009MNRAS.392..945V} {392, 945}

\bibitem[\protect\citeauthoryear{{Vegetti}, {Czoske}  \& {Koopmans}}{{Vegetti}
  et~al.}{2010a}]{Vegetti_2010_1}
{Vegetti} S.,  {Czoske} O.,   {Koopmans} L.~V.~E.,  2010a, \mn@doi [\mnras]
  {10.1111/j.1365-2966.2010.16952.x}, \href
  {http://adsabs.harvard.edu/abs/2010MNRAS.407..225V} {407, 225}

\bibitem[\protect\citeauthoryear{{Vegetti}, {Koopmans}, {Bolton}, {Treu}  \&
  {Gavazzi}}{{Vegetti} et~al.}{2010b}]{Vegetti_2010_2}
{Vegetti} S.,  {Koopmans} L.~V.~E.,  {Bolton} A.,  {Treu} T.,   {Gavazzi} R.,
  2010b, \mn@doi [\mnras] {10.1111/j.1365-2966.2010.16865.x}, \href
  {http://adsabs.harvard.edu/abs/2010MNRAS.408.1969V} {408, 1969}

\bibitem[\protect\citeauthoryear{{Vegetti}, {Lagattuta}, {McKean}, {Auger},
  {Fassnacht}  \& {Koopmans}}{{Vegetti} et~al.}{2012}]{vegetti2012}
{Vegetti} S.,  {Lagattuta} D.~J.,  {McKean} J.~P.,  {Auger} M.~W.,  {Fassnacht}
  C.~D.,   {Koopmans} L.~V.~E.,  2012, \mn@doi [Nature] {10.1038/nature10669},
  \href {http://adsabs.harvard.edu/abs/2012Natur.481..341V} {481, 341}

\bibitem[\protect\citeauthoryear{{Vegetti}, {Koopmans}, {Auger}, {Treu}  \&
  {Bolton}}{{Vegetti} et~al.}{2014}]{2014MNRAS.442.2017V}
{Vegetti} S.,  {Koopmans} L.~V.~E.,  {Auger} M.~W.,  {Treu} T.,   {Bolton}
  A.~S.,  2014, \mn@doi [\mnras] {10.1093/mnras/stu943}, \href
  {http://adsabs.harvard.edu/abs/2014MNRAS.442.2017V} {442, 2017}

\bibitem[\protect\citeauthoryear{Yèche, Palanque-Delabrouille, Baur  \&
  du~Mas~des Bourboux}{Yèche et~al.}{2017}]{Yeche:2017upn}
Yèche C.,  Palanque-Delabrouille N.,  Baur J.,   du~Mas~des Bourboux H.,
  2017, \mn@doi [JCAP] {10.1088/1475-7516/2017/06/047}, 1706, 047

\makeatother
\end{thebibliography}
\label{lastpage}
\end{document}